\documentclass[conference]{IEEEtran}
\IEEEoverridecommandlockouts
\usepackage{cite}
\usepackage{amsmath,amssymb,amsfonts}
\usepackage{algorithm}
\usepackage{algpseudocode}
\usepackage{graphicx}
\usepackage{textcomp}
\usepackage{xcolor}
\usepackage{graphicx}  
\usepackage{cite}
\usepackage{svg}      
\usepackage{caption}   
\usepackage{array} 
\usepackage{stfloats}
\usepackage{tabularx}
\usepackage{amsmath}
\usepackage{changepage}
\usepackage[hidelinks]{hyperref}
\usepackage{float}
\usepackage{soul}
\usepackage{subcaption}
\usepackage{booktabs}
\def\BibTeX{{\rm B\kern-.05em{\sc i\kern-.025em b}\kern-.08em
    T\kern-.1667em\lower.7ex\hbox{E}\kern-.125emX}}
\newcommand{\nomenentry}[2]{%
  \noindent\hangindent=0.3em\hangafter=0%
  \makebox[3.9em][l]{\hspace{0.2em}$#1$}
  #2\par\vspace{0.6ex}
}

\begin{document}

\title{Reinforcement Learning versus Optimization for Optimal Transmission Switching: \\A Comparative Study \\
}

\author{
\IEEEauthorblockN{Israel Abiala and Yuanrui Sang}
\IEEEauthorblockA{\textit{Dept. of Electrical and Computer Engineering} \\
\textit{University of Massachusetts Amherst}\\
Amherst, MA, USA \\
\{iabiala,ysang\}@umass.edu}
\and
\IEEEauthorblockN{Rachel (Ryan) M. Gerdes}
\IEEEauthorblockA{\textit{Dept. of Electrical and Computer Engineering} \\
\textit{Virginia Tech}\\Arlington, VA, United States \\
rgerdes@vt.edu}
}

\maketitle
\begin{abstract}
Optimal Transmission Switching (OTS) reduces generation cost by strategically opening transmission lines, but its mixed-integer linear program (MILP) formulation scales poorly for large-scale transmission networks. Reinforcement learning (RL) offers a computationally efficient alternative, but existing RL-based OTS approaches rely on soft penalties that permit physical constraint violations. This paper presents a comparison between an RL framework and an MILP-based optimization method for OTS. Case studies were carried out on the IEEE RTS-96 24-bus system; results show that the agent was able to produce near-optimal solutions at low switching budgets and tended to yield suboptimal solutions at high switching
budgets. However, the RL agent was able to generate feasible solutions two-to-three orders of magnitude faster than the optimization solver.

\end{abstract}


\section*{Nomenclature}

\textbf{Sets/Indices} \\

\nomenentry{\mathbb{B}}{Set of buses}%
\nomenentry{\mathbb{G}}{Set of generators}%
\nomenentry{\mathbb{G}_n}{Set of generators connected to bus $n$}%
\nomenentry{\mathbb{L}}{Set of transmission lines}%
\nomenentry{\mathbb{S}}{Set of linearized generation segments}%
\nomenentry{\mathbb{\delta}^{in}_n}{Set of lines entering bus n}%
\nomenentry{\mathbb{\delta}^{out}_n}{Set of lines leaving bus n}%

\vspace{1ex}
\textbf{Parameters} \\

\nomenentry{C^{nl}_g}{No load cost of generator $g$}%
\nomenentry{C^{seg}_{g,s}}{Marginal cost of segment $s$ of generator $g$}%
\nomenentry{F^{min}_k}{Minimum thermal capacity of line $k$}%
\nomenentry{F^{max}_k}{Maximum thermal capacity of line $k$}%
\nomenentry{S_{base}}{Base MVA of the system}%
\nomenentry{x_\ell}{Reactance of line $k$ }%
\nomenentry{b_\ell}{Electrical susceptance of line $k$ ($= S_{base}/x_k$)}%
\nomenentry{P_g^{min}}{Minimum generation of generator $g$}%
\nomenentry{P_g^{max}}{Maximum generation of generator $g$}%
\nomenentry{d_n}{Demand at bus $n$}%
\nomenentry{M}{Large number}%
\nomenentry{C}{Upper limit on the number of open transmission lines}%
\nomenentry{P^{seg}_{g,s}}{Upper limit of segment $s$ of generator $g$}%
\nomenentry{\theta^{min}_n}{Minimum voltage angle of bus $n$}%
\nomenentry{\theta^{max}_n}{Maximum voltage angle of bus $n$}%

\vspace{1ex}
\textbf{Variables} \\

\nomenentry{z_k}{Binary variable indicating whether line $k$ is open\\
or close}%
\nomenentry{D_{n}}{Load demand at bus $n$}%
\nomenentry{F_{k}}{Active power flow of line $k$}%
\nomenentry{\theta_n}{Voltage angle of bus $n$}%
\nomenentry{P^{seg}_{g,s}}{Power generation of segment $s$ of generator $g$}
\nomenentry{P_{g}}{Total generation by generator $g$}%

\section{Introduction}

As transmission congestion increases as a result of the integration of renewable energy sources into the power grid, transmission switching is used to mitigate the congestion and reduce power system operating costs \cite{numan2023role}. According to the National Transmission Needs Study report released in 2023, implementing grid enhancing technologies like transmission switching can increase the effective utilization of transmission lines by 16\% and reduce line overloading by at least 40\% \cite{doe2023national}.

Several studies have investigated the optimal transmission switching (OTS) problem for different test cases. Traditionally, it is formulated as a DC optimal power flow (DCOPF)-based optimization model, which determines the optimal network topology and generator dispatch to meet a certain demand. In this case, the optimization problem is a mixed Integer problem, which is NP-hard but can be effectively solved for small-scale systems and significantly reduce transmission congestion, and, consequently, generation dispatch cost \cite{fisher2008optimal}. To study the long-term impact of transmission switching, the OTS problem can be integrated into unit commitment models considering N-1 security \cite{5401077}.  
The OTS problem can also be integrated into a long-term planning model, so that transmission expansion can be done considering the flexibility that transmission switching can offer \cite{5409537}. In the event of a transmission outage, the network topology can be adjusted to enhance transmission transfer capability and reduce power outages. An extensive formulation was developed in \cite{7540982} to reduce outages through corrective switching of transmission lines. The impact on the electricity customers can be reduced when optimal lines are switched out during maintenance or unplanned outages. 

Different algorithms have been adopted over the years to conduct OTS. To enhance the computational efficiency of OTS problems, a method was proposed to calculate the selection and impact of big-M values \cite{10012185} in the MILP model. This method was tested on the IEEE RTS-96 test system, and an optimal big-M value was found that reduces the computational time while minimizing the costs. A bound-strengthening method was implemented on the standard OTS problem to reduce computational time, making the model scalable for large-scale power systems \cite{8451930}. A three-stage robust MILP model was developed to manage OTS with renewable energy and n-k contingencies. In its solution approach, it combined a nested column-and-constraint generation algorithm and the Dantzig-Wolfe procedure \cite{10129825}. In addition, parallel heuristics can be used to speed up the solution process \cite{9857996}.

Due to the computational complexity associated with the OTS problem, machine learning has been increasingly adopted to 
address its tractability issues. This evolution has progressed from supervised learning 
toward modern deep reinforcement learning strategies capable of 
autonomously making dispatch and line-switching decisions under 
varying load scenarios. A two-stage architecture is proposed in 
\cite{meng2025flow} that uses a line-graph neural network (LGNN) 
to compute DC power flows and a separate graph neural network to switch lines via circuit breakers. 
Tang \textit{et al.}~\cite{tang2022optimal} applies a Deep Q-Network 
(DQN)-based RL algorithm to reduce short-circuit currents on the 
IEEE 30-bus and 118-bus systems, demonstrating that the agent can 
select effective branches at each time step. A physics-informed variant is proposed in 
\cite{dogoulis2026physics}, combining a Proximal Policy 
Optimization (PPO) algorithm with semi-Markov control and a 
Gibbs prior that encodes environmental physics; the method 
improves upon conventional PPO.
In \cite{11023158}, a Mixed-Integer Linear Program (MILP)-based OTS is solved using machine learning methods and parameter optimization to minimize generation costs. These ML models were tested on AC power flow, and it was found to significantly reduce costs by 44\% for some case studies. 
To balance generation cost with feasibility, Lin \textit{et al.}~\cite{lin2025deep} develop a multi-objective transmission switching model using a Q-value network within a discrete Soft Actor-Critic (SAC) framework to predict optimal switching actions, reporting improvements over existing DRL algorithms. A more recent effort explored quantum RL for large-scale power grids~\cite{lin2025two}, achieving faster convergence and more stable training than classical baselines. Lu \textit{et al.}~\cite{lu2024research} apply a two-stage DRL based on Asynchronous Advantage Actor-Critic to corrective transmission switching, reducing the computational burden of traditional methods.

It is difficult for RL-only algorithms to achieve stable convergence and training stability. In some cases, the RL agent may learn an optimal policy during training and later settle for a sub-optimal one due to the entropy used. Safe RL strategies such as action projection can be used to reduce the constraint violations in the RL agent's decisions \cite{eichelbeck2022contingency}. The power grid can be forced into a blackout if the policy proposed by the agent is not projected incrementally while maintaining system balance \cite{zhou2021action}.

In summary, there exists a variety of optimization-based and machine learning-based methods to solve the OTS problem. However, there is still a lack of work that compares the pros and cons of both approaches. Hence, this study aims to compare the RL-based method with the optimization-based method for OTS in terms of solution optimality and computational efficiency. The RL method in this study adopts behavioral cloning to accelerate the agent's discovery of a near-optimal policy. The SAC agent is first pretrained on demonstrations generated by the MILP solver and then fine-tuned with the per-constraint soft-Lagrangian reward, thereby reducing exploration time and improving convergence.

The main contributions of the proposed work are as follows:
\begin{itemize}
  \item A \textbf{per-constraint soft-Lagrangian} formulation of DC optimal transmission switching is proposed: every constraint family (generator limits,
        segment capacity, thermal, DC/Ohm, power balance, angle, reference,
        line-budget, and binary integrality) is written as a non-negative
        violation and priced in the reward by its own multiplier $\lambda_i$.
  \item \textbf{Automatic multiplier tuning via per-constraint dual ascent},
        which raises the price of persistently violated constraints and decays
        satisfied ones; the converged multipliers are then frozen to yield a
        stable fixed-$\lambda$ policy.
  \item A \textbf{per-constraint normalization} scheme that renders the
        multipliers commensurate---each violation scaled by its own
        characteristic range, with per-line \emph{fractional} overload for
        thermal/DC and a small balance reference so power balance binds---together
        with violation clipping for training stability.
  \item \textbf{Behavioral cloning from MILP-OTS demonstrations} to warm-start a
        \emph{single unified} policy conditioned on (load factor, switching
        budget), covering all budgets $j\in\{0,\dots,5\}$ with one network.
    \item \textbf{A comparison} of the proposed RL method with the traditional optimization-based method is performed.
\end{itemize}

The rest of the paper is organized as follows.
\hyperref[sec:model]{Section~II} discusses the model formulation.
\hyperref[sec:setup]{Section~III} focuses on the simulation setup.
\hyperref[sec:analysis]{Section~IV} describes results analysis and taking considering one load factor.
Lastly, \hyperref[sec:conclusion]{Section~V} concludes the paper by providing key insights, and suggests some possible future work.

\section{Model Formulation}

We formulate the OTS problem in two 
ways: (i) a mixed-integer linear program (MILP) serving as the 
optimality benchmark, and (ii) a Markov Decision Process (MDP) 
describing the power grid environment in which the RL agent 
interacts. Both formulations build on the DCOPF
model subject to the physical constraints of the system.

\subsection{Optimization Formulation}
The optimization-based method \cite{fisher2008optimal} is the foundation for the RL-based method. In the optimization-based model, the objective function aims at minimizing the total operational cost, which is shown in \eqref{eq:objective}, which includes:
\begin{itemize}
    \item No-load cost of each generator, and
    \item Generation cost calculated from a piecewise linear cost function.
\end{itemize}

\begin{equation}
\min_{} 
\sum_{g=1}^{G}\left[c_g^{nl} + \sum_{seg=1}^{N_{seg}} c_{g,seg} \, P_{g,seg}\right]
\label{eq:objective}
\end{equation}

This optimization problem is subject to a number of constraints:
\begin{align}
P_g^{min} \leq P_g \leq P_g^{max}, \quad\quad &\forall g \in \mathbb{G} \label{eq:gen_bounds}\\
P_g = \sum_{s \in \mathbb{S}} \mathbf{p}_{g,s}^{seg}, \quad\quad\quad\quad &\forall g \in \mathbb{G} \label{eq:seg_sum}\\
0 \leq \mathbf{p}_{g,s}^{seg} \leq P_{g,s}^{seg}, \quad\quad\quad &\forall g \in \mathbb{G}, \, s \in \mathbb{S} \label{eq:seg_bounds}\\
F_k^{min} z_k \leq F_k \leq F_k^{max} z_k, \quad &\forall k \in \mathbb{L} \label{eq:thermal}\\
-b_k(\theta_{f(k)} - \theta_{t(k)}) - F_k \leq (1-z_k) M, \; &\forall k \in \mathbb{L} \label{eq:bigM_ub}\\
-b_k(\theta_{f(k)} - \theta_{t(k)}) - F_k \geq -(1-z_k) M, \; &\forall k \in \mathbb{L} \label{eq:bigM_lb}\\
\sum_{g \in \mathbb{G}_n} \!\! P_g \, - \! \sum_{k \in \delta_n^{out}} \!\! F_k + \! \sum_{k \in \delta_n^{in}} \!\! F_k = D_n, \; &\forall n \in \mathbb{B} \label{eq:balance}\\
\theta_n^{min} \leq \theta_n \leq \theta_n^{max}, \quad\quad &\forall n \in \mathbb{B} \label{eq:theta_bounds}\\
\theta_0 = 0 \quad\quad\quad\quad\quad&\label{eq:slack}\\
\sum_{k \in \mathbb{L}} (1 - z_k) \leq C \quad\quad\quad&\label{eq:gub}\\
z_k \in \{0, 1\}, \quad\quad\quad\quad &\forall k \in \mathbb{L} \label{eq:binary}
\end{align}

The generator upper and lower limits are enforced by \eqref{eq:gen_bounds}--\eqref{eq:seg_bounds}, which also ensure that the total generator output equals the sum of its four cost segments. Constraint \eqref{eq:thermal} imposes the thermal capacity limit on each line $k$, which is activated or deactivated by the binary switching variable in \eqref{eq:binary}. Constraints \eqref{eq:bigM_ub}--\eqref{eq:bigM_lb} enforce DC power flow and disable power flow on open lines through the big-M method, where $M$ is chosen sufficiently large to render the coupling inactive when $z_k = 0$. Constraint \eqref{eq:balance} enforces nodal power balance by equating net generator injection and net line flow at each bus $n \in \mathbb{B}$. Voltage angles are bounded by \eqref{eq:theta_bounds} with the slack bus fixed at $\theta_0 = 0$ via \eqref{eq:slack}, and \eqref{eq:gub} limits the total number of simultaneously open lines, i.e., the switching budget, to $C$, and \eqref{eq:binary} enforces that each line status $z_k$ is a binary decision ($z_k = 1$ for closed, $z_k = 0$ for open).

\subsection{Reinforcement Learning Formulation}
\label{subsec:mdp}

The OTS problem \eqref{eq:objective}--\eqref{eq:binary} is NP-hard, with
solution time growing exponentially in $|\mathbb{L}|$, making it hard to scale for large-scale systems during real-time operation. We reformulate OTS as a single-step MDP
$\mathcal{M} = (\mathcal{S}, \mathcal{A}, r)$ with $\gamma = 0$ (one decision per
episode). Each episode is one operating snapshot with load factor
$\ell$ scaling nodal demand as
$D_n = \ell\, D_n^{\text{peak}}$.

\subsubsection{Observation Space}
A \emph{single unified} policy is conditioned on the load level, the switching
budget, and a base-case congestion map, so one network serves all budgets:
\begin{equation}
\mathbf{s}_t = \big[\,\ell_t,\; j_t,\; \{f_{k,t}\}_{k \in \mathbb{L}}\,\big],
\qquad
f_{k,t} = \min\!\Big(\tfrac{|F_k^{\text{base}}(\ell_t)|}{F_k^{\max}},\,\kappa_\ell\Big),
\label{eq:obs}
\end{equation}
where $j_t \in \{0, 1, 2, \dots,C\}$ is the number of lines permitted open during this episode,
and $f_{k,t}$ is the per-line loading of the \emph{all-lines-in} base case at
load factor $\ell_t$. The loadings expose which lines are congested ($f_{k,t}\!\sim\!1$),
which helps guide the decisions for transmission switching. The
observation dimension is $2+|\mathbb{L}|$, making it 40 observation spaces.

\subsubsection{Action Space}
The continuous action outputs raw dispatch and raw switching signals:
\begin{equation}
\mathbf{a}_t = \big[\{P_g\}_{g \in \mathbb{G}},\; \{z_k\}_{k \in \mathbb{L}}\big],
\label{eq:action}
\end{equation}
with $P_g \in [P_g^{\min}, P_g^{\text{high}}]$ and $z_k \in [0,1]$. 
Then the power flow is calculated by $F_k = -b_k\,z_k\,(\theta_{f(k)}-\theta_{t(k)})$, 
and the thermal limit constraint violations are penalized in the reward function.
Eventually, a line is reported open when $z_k < 0.5$. 

% Both are used
% \emph{directly}: dispatch is not projected, and $z_k$ enters the DC power flow as
% a \emph{fractional} susceptance $b_k z_k$ (no binarization). The binary and
% line-budget penalties drive $z_k \to \{0,1\}$ and toward exactly $C=j$ open lines;
% . Given $(P,z)$, the DC solve recovers bus
% angles $\theta$ and flows $F = b\,z\,(\theta_{\text{from}}-\theta_{\text{to}})$,
% so the Ohm and reference-bus constraints hold by construction.

\subsubsection{Reward Function}
The reward is the negative normalized generation cost minus the per-constraint
soft-Lagrangian penalty:
\begin{equation}
r_t = -\frac{J^{R}(P_t)}{C_{\max}} \;-\; \sum_{i}\lambda_i\,\hat g_i,
\qquad
\hat g_i = \min\!\Big(\frac{g_i}{s_i},\, \kappa\Big),
\label{eq:reward}
\end{equation}
where $J^{R}$ is the piecewise-linear generation cost, $C_{\max}$ the full-output
cost, $g_i \ge 0$ the violation of constraint family $i$, $s_i$ its normalizer, $\kappa=5$ the violation
clip, and $\lambda_i \ge 0$ the per-constraint multiplier. The $\lambda_i$ are
obtained online by projected dual ascent,
$\lambda_i \leftarrow \big[\lambda_i + \eta\,(\overline{\hat g_i}-\tau_i)\big]_0^{\lambda_{\max}}$.

\subsubsection{Training Algorithm}
The policy $\pi_\phi$ is a Gaussian actor with hidden layers of $256 \times 256$, and the activation function used was the Exponential Linear Unit (ELU)
and twin critics, $Q_{\psi_1}$ and $Q_{\psi_2}$, trained with SAC~\cite{haarnoja2018soft} augmented by behavioral cloning.
% the actor minimizes the SAC objective plus $\lambda_{\text{BC}}$ times a behavioral-cloning
% loss on MILP demonstrations. 
The demonstrations are held in a permanent buffer
$\mathcal{D}_{\text{demo}}$ and sampled into every gradient batch alongside the
environment replay buffer $\mathcal{D}$.

\section{Simulation Setup}
\label{sec:setup}

We evaluate the proposed framework on the IEEE RTS-96 system with
$24$ buses, $38$ transmission lines, and
$32$ generators. Each generator's cost is a four-segment
piecewise-linear function plus a no-load cost. Each episode samples a load factor uniformly at random,
$\ell \sim \mathcal{U}(0.5487,\,0.93)$. Evaluation uses the deterministic
(mean-action) policy on $100$ load factors drawn uniformly at random from the
same range, for every switching budget $j \in \{0,1,\ldots,5\}$.

A \emph{single unified policy} is trained, conditioned on the observation: the environment resamples the budget $j$ each reset, so one network
covers all switching budgets rather than a separate policy per budget.
Training proceeds in two phases using a Gymnasium environment. In
\textbf{behavioral cloning (BC)}, one MILP-OTS solution per $(\ell, j)$ pair
(over all six budgets) pre-trains the actor via maximum likelihood. In
\textbf{SAC fine-tuning}, the policy is trained with SAC
\cite{haarnoja2018soft} under the per-constraint soft-Lagrangian reward, with BC
regularization retained throughout ($\lambda_{\text{BC}}=0.4$). Table
\ref{tab:hyper} lists the hyperparameters.

\begin{table}[htbp]
\caption{Training Hyperparameters}
\begin{center}
\begin{tabular}{|l|l|}
\hline
\textbf{Parameter} & \textbf{Value} \\
\hline
\multicolumn{2}{|c|}{\textit{Behavioral Cloning}} \\
\hline
MILP demos & one per $(\ell, j)$ pair \\
\hline
Epochs & 3{,}000 \\
\hline
Optimizer / loss & Adam / max-likelihood (NLL) \\
\hline
Learning rate & $5 \times 10^{-4}$ \\
\hline
\multicolumn{2}{|c|}{\textit{Soft Actor-Critic}} \\
\hline
Discount factor $\gamma$ & 0 (single-step) \\
\hline
Learning rate (actor \& critic) & $3 \times 10^{-4}$ \\
\hline
Replay buffer size & $4 \times 10^{5}$ \\
\hline
Batch size & 500 \\
\hline
Hidden layers (actor \& critic) & $[256, 256]$, ELU \\
\hline
Target entropy $\mathcal{H}_{\text{target}}$ & $-70$ \\
\hline
Entropy coefficient $\alpha$ & auto-tuned \\
\hline
Warm-up steps (\texttt{learning\_starts}) & 10{,}000 \\
\hline
Train frequency / gradient steps & 500 / 1{,}000 \\
\hline
Total training steps & 100{,}000 \\
\hline
BC regularization weight $\lambda_{\text{BC}}$ & 0.4 \\
\hline
\multicolumn{2}{|c|}{\textit{Soft Lagrangian}} \\
\hline
Multiplier mode & fixed (frozen from dual ascent) \\
\hline
Violation clip $\kappa$ & 5.0 \\
\hline
Balance normaliser $s_{\text{kcl}}$ & 100 MW \\
\hline
Balance $\varepsilon$-deadband (annealed) & $10 \to 0.5$ MW \\
\hline
\end{tabular}
\label{tab:hyper}
\end{center}
\end{table}

We evaluate the unified RL policy across 6 switching budgets on 100 independent operating snapshots per budget. For every snapshot, the policy produces, in a single snapshot, both the generator dispatch and the switching signal.

\section{Results and Analysis}
\label{sec:results}

The RL results are compared against those obtained by solving the MILP-based OTS problem, defined by Equations \eqref{eq:objective}--\eqref{eq:binary}, via Python PuLP with the HiGHS solver. Evaluation metrics include generation cost savings vs.\ the no-switching case, line-match between the RL and MILP open-line sets (exact-set match and the per-line fraction recovered), and computational efficiency. All experiments used random seed~42.
 %We benchmark each RL solution against the MILP-optimal solution of the same instance \eqref{eq:objective}--\eqref{eq:binary}, solved to optimality. 

\subsection{Cost Savings}
\begin{figure}[t]
  \centering
  \includegraphics[width=\linewidth]{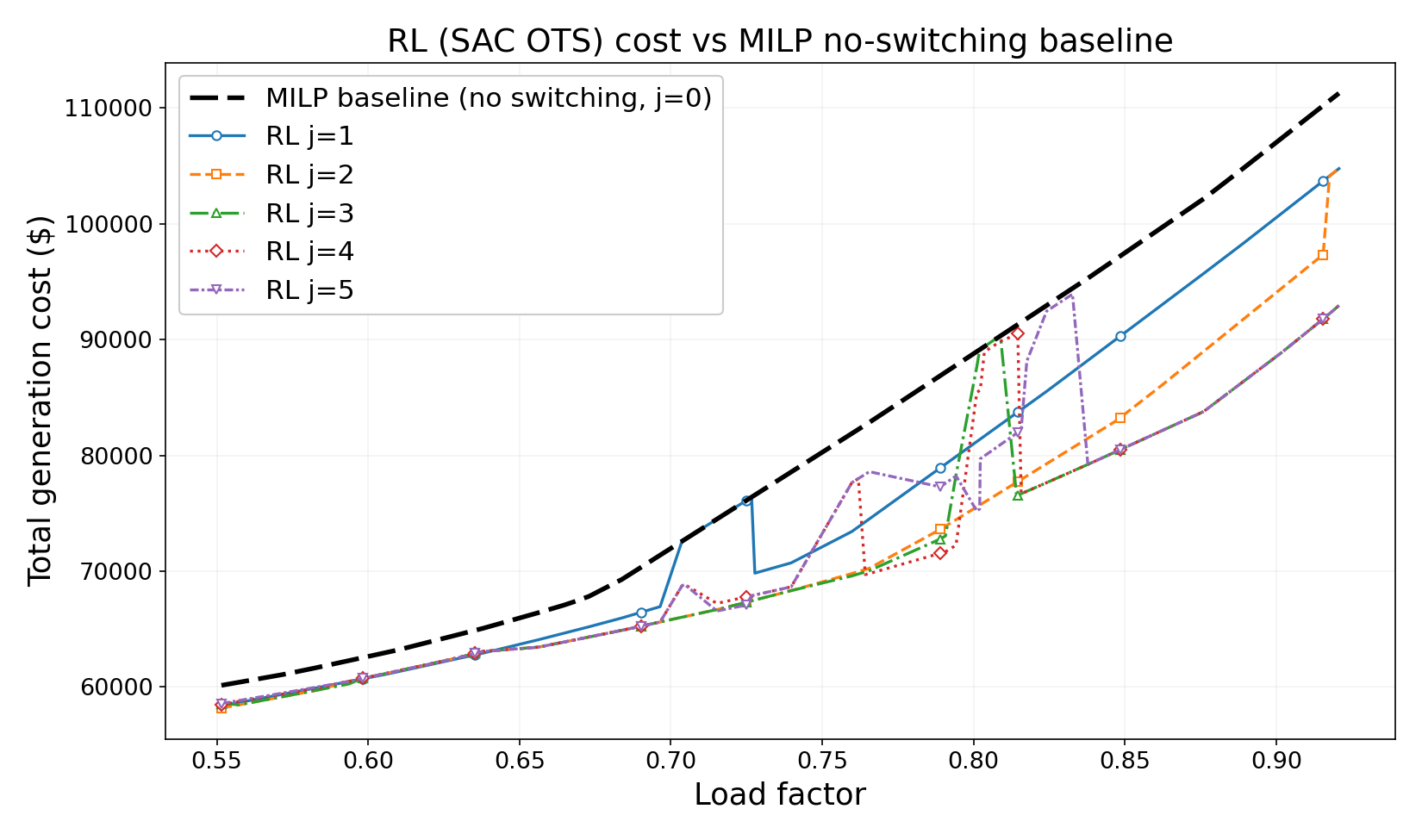}
  \caption{Total generation cost of the RL (SAC-OTS) policy %under switching budgets $j=1,\dots,5$ versus the no-switching MILP baseline ($j{=}0$, dashed)%, plotted against the system load factor. %The RL curves fall increasingly below the baseline as load rises, illustrating that optimal transmission switching yields its largest savings under heavy congestion.
  }
  \label{fig:savings}
\end{figure}

\begin{figure}[t]
  \centering
  \includegraphics[width=\linewidth]{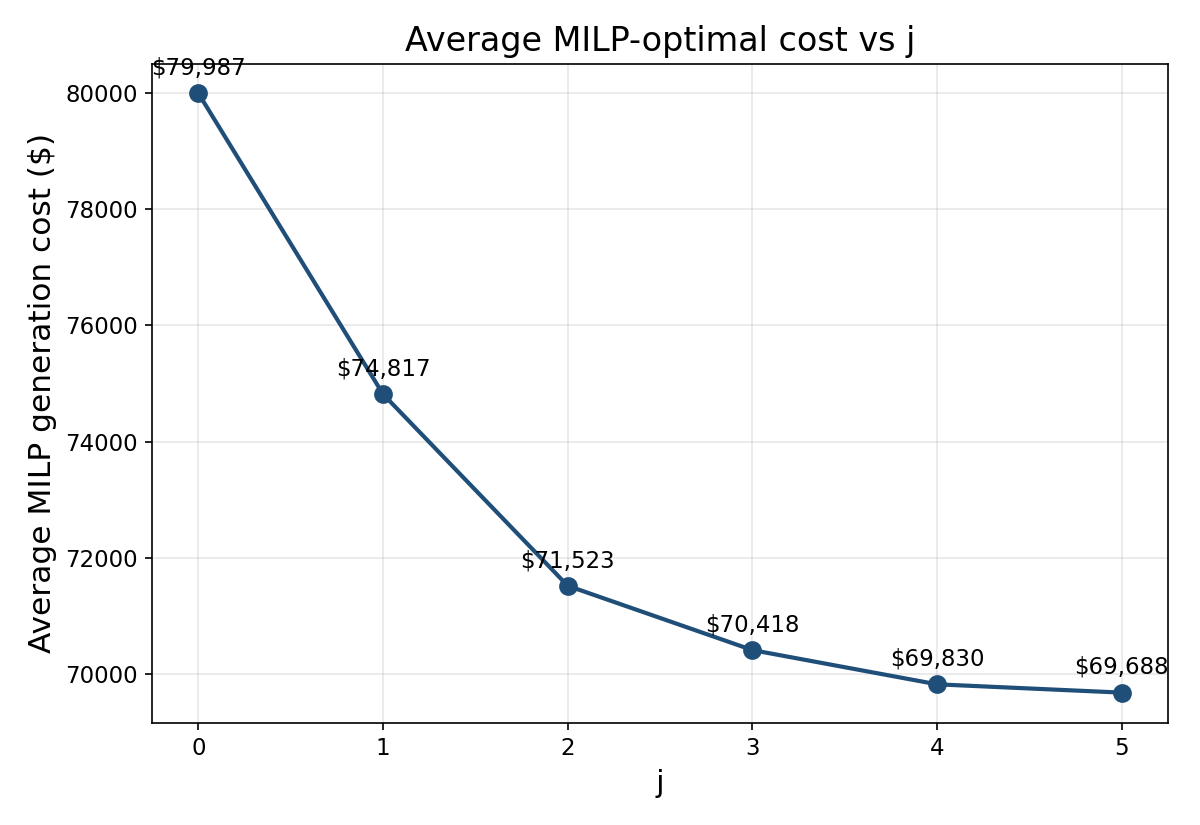}
  \caption{MILP average cost savings across each j for the 100 load factors.
  }
  \label{fig:avgsavings}
\end{figure}

Fig.~\ref{fig:savings} plots the total generation cost of the RL policy under
each switching budget against the system load factor, with the no-switching MILP
baseline ($C{=}0$) shown dashed. Every RL curve lies on or below the baseline,
confirming that the learned switching actions never increased cost and that the
policy correctly defaulted to the no-switching solution when switching offers no
benefit. At low load factors ($\ell \lesssim 0.65$) the transmission system was less congested, the
curves for all budgets nearly coincide with one another, and the mean savings were small ($\sim\!\$1.8$k per scenario). As load increased the
RL curves separate from the baseline and the gaps widened, reaching
about $\sim\!\$12.2$k per scenario for $\ell > 0.80$. The economic value of OTS concentrated in the heavily congested systems, where the policy delivered its largest cost savings.

Aggregated over all 100 load factors, the mean saving relative to the
no-switching baseline rose from $\sim\!\$4.5$k at $C{=}1$ to $\sim\!\$8.2$k
at $C{=}2$ and $\sim\!\$8.5$k at $C{=}3$, then saturated near $\$7.9$k for
$C{=}4$ and $C{=}5$. Measured against the MILP-optimal savings at each budget, the policy captured $88\%$, $97\%$ and $89\%$ of the attainable benefit with a switching budget of 1, 2, and 3, respectively. The percentage of attained benefits reduced to $78\%$ and $77\%$ for switching budgets of 4 and 5, respectively. As the optimal solution increasingly demanded more coordinated switches, the coordination became increasingly difficult for the RL agent. Also, the solutions obtained from the HiGHS solver showed that the savings started to saturate when the switched line reached 3, and additional switches yield only marginal incremental savings, as Fig. \ref{fig:avgsavings} shows. This shows that there is a trade-off between the incremental benefits and RL effectiveness when deciding the number of lines to open. 
% =====================================================================\subsection{Switching Decisions}
\subsection{Switching Decisions}
\label{subsec:lines}

\begin{figure}[t]
  \centering
  \begin{subfigure}[t]{\linewidth}
    \includegraphics[width=\linewidth]{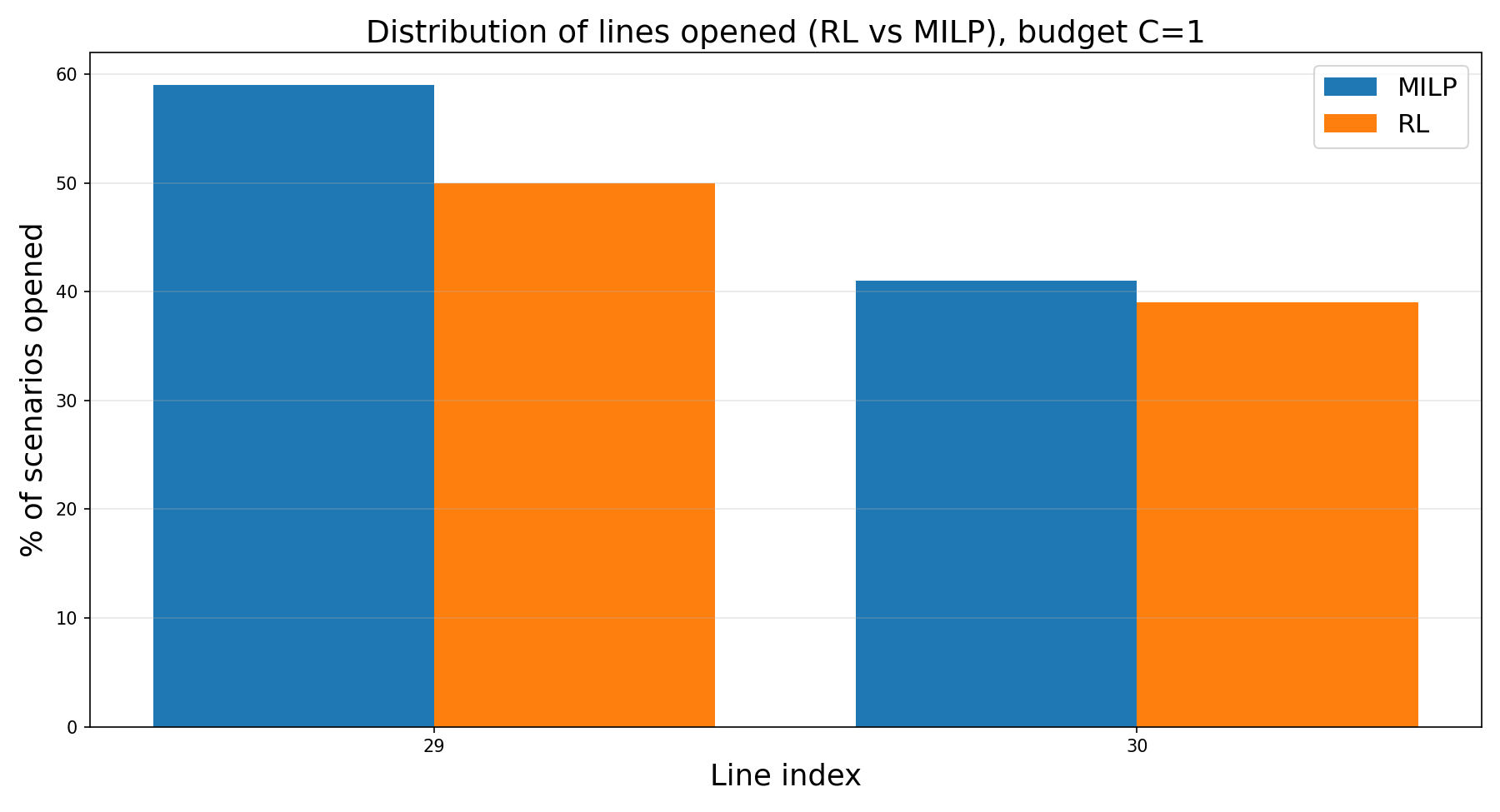}
    \caption{$C=1$}\label{fig:lines_c1}
  \end{subfigure}\hfill
  \begin{subfigure}[t]{\linewidth}
    \includegraphics[width=\linewidth]{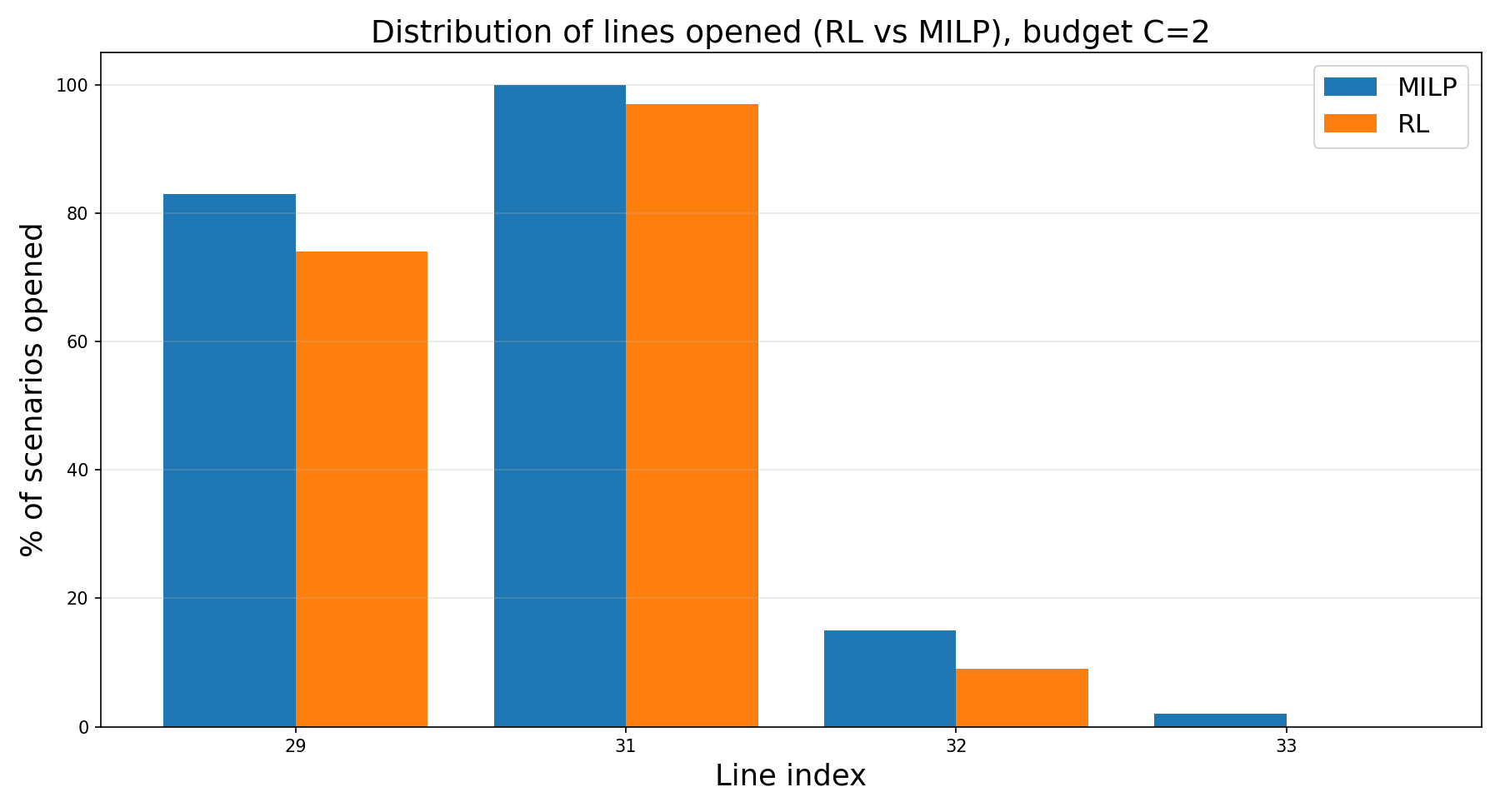}
    \caption{$C=2$}\label{fig:lines_c2}
  \end{subfigure}
  \caption{Lines opened (RL vs.\ MILP) at low budgets.}
\end{figure}

\begin{figure}[ht]
  \centering
  \includegraphics[width=\linewidth]{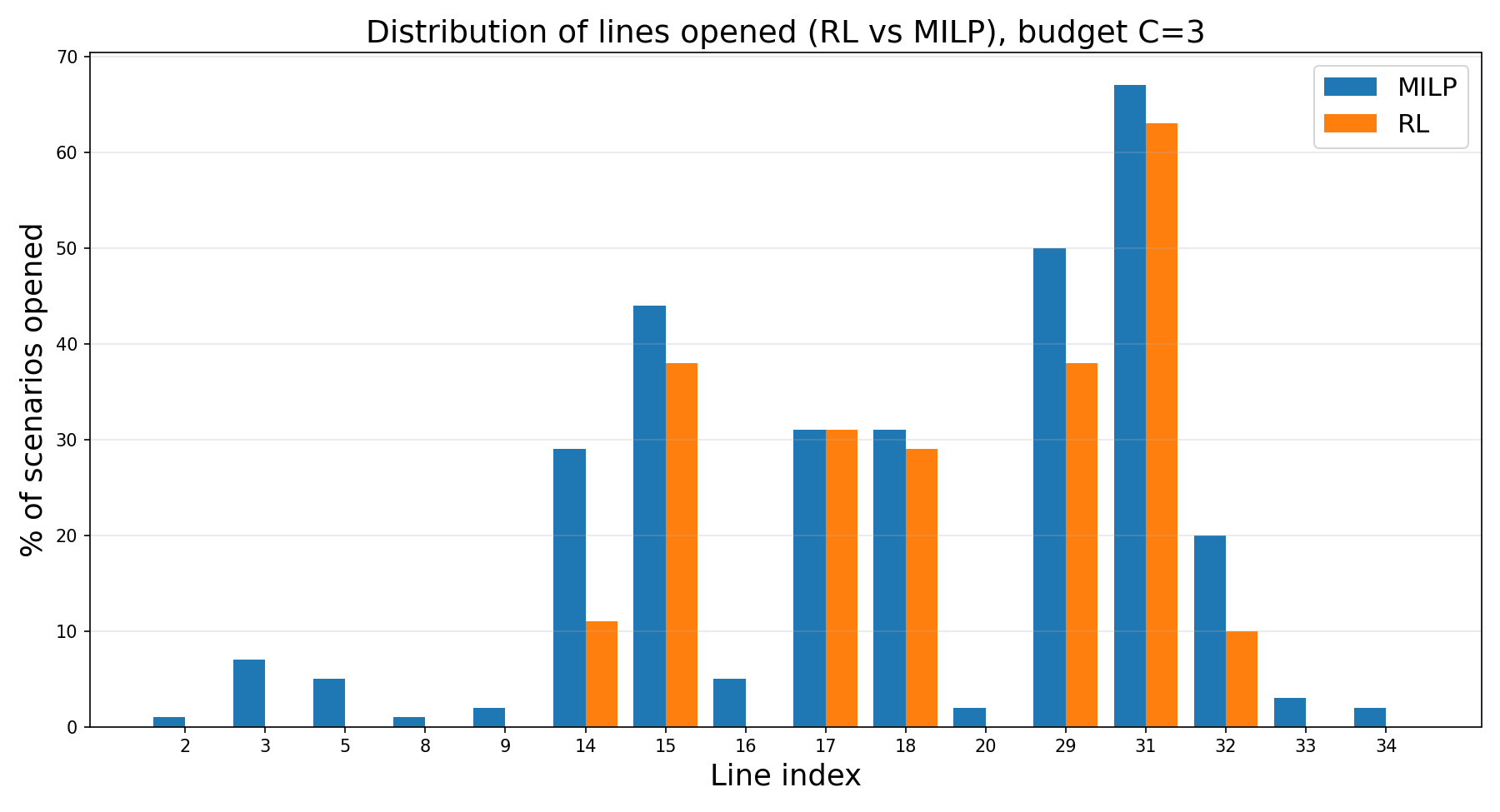}
  \caption{Lines opened (RL vs.\ MILP) at $C=3$.}\label{fig:lines_c3}
\end{figure}

\begin{figure}[ht]
  \centering
  \begin{subfigure}[t]{\linewidth}
    \includegraphics[width=\linewidth]{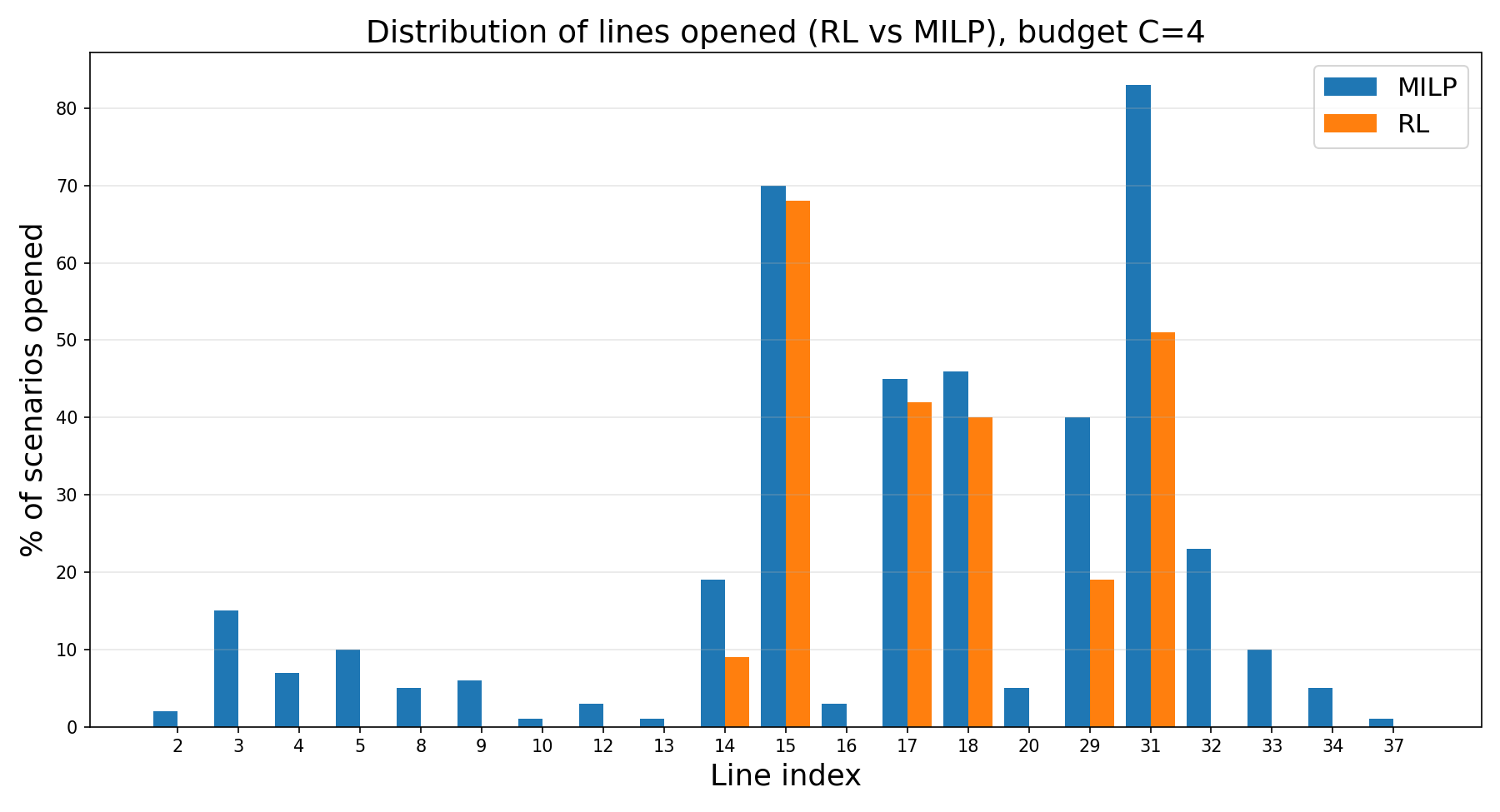}
    \caption{$C=4$}\label{fig:lines_c4}
  \end{subfigure}\hfill
  \begin{subfigure}[t]{\linewidth}
    \includegraphics[width=\linewidth]{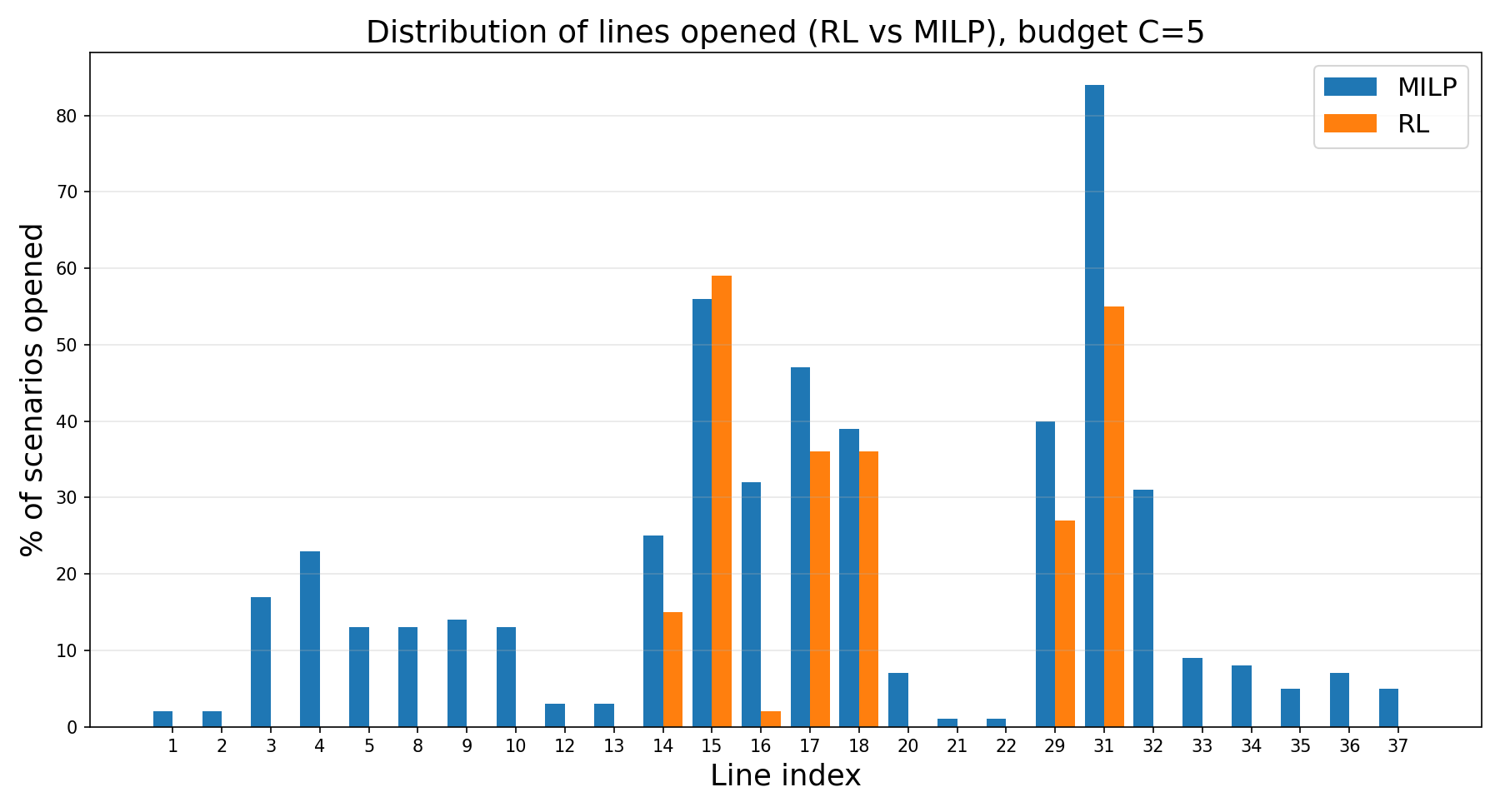}
    \caption{$C=5$}\label{fig:lines_c5}
  \end{subfigure}
  \caption{Lines opened (RL vs.\ MILP) at high budgets.}
\end{figure}

For each switching budget, we report the fraction of scenarios in which every line is opened
by RL and by MILP (Fig.~\ref{fig:lines_c1}--\ref{fig:lines_c5}). Three regimes
emerge.

\textbf{Low budget ($C{=}1,2$): RL almost reproduced the optimal distribution.} At
$C{=}1$, as Fig.~\ref{fig:lines_c1} shows, the optimum is a load-dependent choice between
two lines, L29 ($\sim\!59\%$ of the cases) and L30 ($\sim\!41\%$ of the cases). The switches from RL matches with the optimal switching with a high percentage (L29: $\sim\!50\%$ of the cases, L30: $\sim\!39\%$ of the cases, no line switched: $\sim\!11\%$ of the cases). At $C{=}2$,
as Fig.~\ref{fig:lines_c2} shows, the optimal switched lines are dominated by the pair $\{$L29, L31$\}$. MILP opens L31 in all cases and L29 in $83\%$ of cases, and RL tracks both closely (L31: $\sim\!97\%$, L29: $\sim\!74\%$). Here the results from the RL method are comparable to those from the optimization-based method.

\textbf{Moderate budget ($C{=}3$): the core was kept, but rare lines were dropped.}
MILP's distribution (Fig.~\ref{fig:lines_c3}) broadens to a dominant core
$\{$L14, L15, L17, L18, L29, L31, L32$\}$ plus a low-frequency tail $\{$L2, L3, L5, L8, L9,
L16, L20, L33, L34$\}$. RL reproduced the
core at comparable frequencies but largely omitted the rare tail.

\textbf{High budget ($C{=}4,5$): RL tended to yield suboptimal solutions.}
At $C{=}4,5$ (Figs.~\ref{fig:lines_c4},~\ref{fig:lines_c5}) the MILP optimum
spreaded across more than twenty distinct lines, most appearing in only a few
percent of scenarios. At high budgets many near-optimal topologies existed and
the optimal choice became highly load-specific. RL instead concentrated on a
small, stable core $\{$L15, L17, L18, L31$\}$ (with secondary L14, L29) and opened them at consistently lower frequencies than MILP. In addition, RL tended to open fewer lines than desired at such budgets. For example, RL opened $\sim\!2$--$3$ lines on average at $C{=}5$.

In summary, the agent learned a robust, feasible policy centered around the lines that are \emph{recurrently} beneficial across loads (the L29--L31 and L14--L18 groups) -- exactly the lines MILP opened most often. What it did not
learn consistently was MILP's load-specific tail at high budgets. Crucially, the recurrently-opened core
carried the bulk of the economic value, so dropping
the marginal tail costs the agent more in \emph{topology match} than in
\emph{cost}, leading to feasible but suboptimal solutions at high switching budgets.

\subsection{Computational Efficiency}

The case studies were performed on a Dell Optiplex Tower with 64 GB of RAM and an Intel Core i9 processor with 36 MB of cache, 24 cores, 32 threads, and up to 5.4 GHz turbo. The full pipeline---MILP demonstration generation, behavioral-cloning pretraining, and SAC fine-tuning of a \emph{single} unified policy conditioned on $(\ell, C)$---completed in \textbf{73.3~min}. Because one neural network served all budgets, training was performed once rather
than per switching budget; the dominant portion of computational time was on generating the MILP demonstrations, which scaled combinatorially with the switching budget. 
Once trained, inference was effectively independent of the switching budget. Each decision required (i) constructing the observation---a base-case DC power-flow solve over the all lines in service to obtain the per-line loadings that form the congestion map, followed by (ii) a single forward pass through the actor that returns the dispatch and switching signals. Together these average only
$\sim\!1.4$~ms per decision across all levels (Table~\ref{tab:comptime}).

\begin{table}[H]
  \centering
  \caption{Average computation time per scenario %as a function of the switching budget $C$, over 100 load-factor scenarios. RL times are actor-plus-inference; MILP times are full solve times for the optimal  transmission switching problem.
  }
  \label{tab:comptime}
  \begin{tabular}{c c c c}
    \toprule
   $C$ & RL inference (ms) & MILP Solution time (ms) & Speedup \\
    \midrule
    0 & 1.23 & 13.0  & 11$\times$  \\
    1 & 1.43 & 357.2 & 249$\times$ \\
    2 & 1.39 & 448.0 & 323$\times$ \\
    3 & 1.40 & 531.8 & 381$\times$ \\
    4 & 1.42 & 641.4 & 451$\times$ \\
    5 & 1.33 & 726.4 & 547$\times$ \\
    \midrule
    Mean & 1.37 & 453.0 & $\sim$330$\times$ \\
    \bottomrule
  \end{tabular}
\end{table}

\section{Conclusions}
\label{sec:conclusion}

In this paper, we formulate optimal transmission switching as a single-step reinforcement learning problem in which every constraint of the model is encoded as a soft, per-constraint Lagrangian penalty in the reward. As the switching budget increases, the number of topology combinations grows rapidly and the optimal solution becomes increasingly load-specific, making it difficult for SAC alone to discover the optimal lines to open. Behavioral-cloning pretraining from MILP demonstrations gave the agent a head start and allowed it to settle on a near-optimal policy. At low-to-moderate switching budgets, the agent closely tracked the optimization-based solution. At higher switching budgets, the optimal topology choices became harder to reproduce, however, the RL agent was still able to produce feasible but suboptimal solutions. Moreover, once trained, the policy produced these decisions roughly two to three orders of magnitude faster than the MILP solver---from $249\times$ at
$C=1$ to $547\times$ at $C=5$---and at a runtime essentially independent of $C$, making it well suited to real-time operation. Future work will focus on improving the accuracy of solving optimal transmission switching with RL-based methods.

\bibliographystyle{IEEEtran}
\bibliography{Ref.bib}

\end{document}